\documentclass[reprint,superscriptaddress,aps,prr]{revtex4-2}
\usepackage{amsmath}
\usepackage{graphicx}
\usepackage{dcolumn}
\usepackage{subfigure}
\usepackage{dsfont}
\usepackage{amssymb}
\usepackage{float}
\usepackage{hyperref}

\begin{document}
	\title{Piecemeal Telescope Array: Exponential Precision with Strong Robustness and
		High Efficiency}
	
	\author{Jian Leng}
	\affiliation{ State Key Laboratory of Low Dimensional Quantum Physics, Department of Physics, \\ Tsinghua University, Beijing 100084, China}
	\author{Yi-Xin Shen}
	\affiliation{ State Key Laboratory of Low Dimensional Quantum Physics, Department of Physics, \\ Tsinghua University, Beijing 100084, China}
	\author{Zhou-Kai Cao}
	\affiliation{ State Key Laboratory of Low Dimensional Quantum Physics, Department of Physics, \\ Tsinghua University, Beijing 100084, China}
	\author{Xiang-Bin Wang}
	\email{ Email: xbwang@mail.tsinghua.edu.cn}
	\affiliation{ State Key Laboratory of Low Dimensional Quantum Physics, Department of Physics, \\ Tsinghua University, Beijing 100084, China}
	\affiliation{ Jinan Institute of Quantum Technology and Jinan branch, Hefei National Laboratory, Jinan, Shandong 250101, China}
	\affiliation{ International Quantum Academy, Shenzhen 518048, China}
	\affiliation{ Frontier Science Center for Quantum Information, Beijing 100193, China}
	
	\begin{abstract}
		Optical telescopes are powerful eyes for terrestrial and astronomical detection. Here we propose a new detection method with high efficiency, strong robustness and super precision, as an enhanced technique for optical telescopes in angular locating. In detail, our method requests only small number of incident single-photons, holds strong fault tolerance to any noise and improves the precision by magnitude orders comparing with current optical telescopes. Given these advantages, our method promises an important progress in remote sensing and astrometry, especially in locating the very dark object.
	\end{abstract}
	
	\maketitle

	\section*{I. Introduction}
	
	The technological development of producing and controlling large lens gives revolutionary breakthrough in optical telescopes \cite{telescopes}. The diameter of current large telescopes can reach several meters and they make crucial contributions in terrestrial and astronomical detection \cite{telescopes}. However, it is quite challenging for producing and controlling even larger lenses that limits the further progress of optical telescopes. Notably, in the specific field of stellar physics, the optical array has a significant advantage comparing with optical telescopes \cite{array}. These arrays can extend the equivalent diameter to be around a kilometer because of the technological development of optical interference since 2001 \cite{glindemann2001light}. Here we propose a method using optical interference that can be qualified in another field of angular locating which is important in remote sensing and astrometry. Since it works in a bit-by-bit behavior, we name it as `piecemeal array'. Our piecemeal array allows fairly large observation errors in the observed data which leads to a strong fault tolerance. This robustness property has included all noise effects, such as channel noise and statistical fluctuation due to small data size. Given this fact, our piecemeal array improves the precision by magnitude orders comparing with current optical telescopes, while it can work with small number of incident single-photons and large channel noise.
	
	\section*{II. Motivation, Difficulty, and Main Idea}
	
	Since 2001 \cite{glindemann2001light}, optical interference for visible light has been successfully demonstrated with the baseline length longer and longer, in the magnitude order of hundreds of meters or even kilometer \cite{array}. The goal of this work is to propose a robust observation achieving exponential angular precision given large initial uncertainty of the angle value of the remote target. In the current single-baseline interference method \cite{gottesman2012longer,khabiboulline2019optical,huang2022imaging,marchese2023large} (for more applications of optical interference, see \cite{zanforlin2022optical,howard2019optimal,lupo2020quantum,huang2021quantum,wang2021superresolution,chen2023astrometry,huang2023ultimate}), the angular value $\theta$ is given by
	\begin{align}
		\theta\approx \frac{\lambda }{2\pi X}\phi \label{phi_theta}
	\end{align}
	where $\phi$ is the observed phase difference of lights from two telescopes of baseline $X$, and $\lambda$ is the wavelength of incident light. Suppose there is an observation error $\Delta\phi=|\phi-\hat\phi|$, where $\hat\phi$ is the experimentally observed value of $\phi$. This gives the precision characterized by uncertainty $\Delta\theta$ of observed angular value
	\begin{align*}
		\Delta\theta= |\theta-\hat\theta|=\frac{\lambda }{2 \pi X}\Delta \phi.
	\end{align*}
	At a first look, the angular precision in uncertainty here is proportional to $1/X$. However, the baseline length $X$ cannot be arbitrary value, but in principle has to satisfy the following condition
	\begin{align}
		0\le  X \theta < \lambda\label{condition}
	\end{align}
	to make sure that $\phi$ is in the range of $[0,2\pi)$ according to Eq. (\ref{phi_theta}) so that $\phi$ is observable in the experiment. Given the initial knowledge $0\le \theta < \bar{\theta}$, Eq. (\ref{condition}) shows that  the largest value of $X$ is $\lambda/\bar{\theta}$ that gives
	\begin{align}
		\Delta\theta =\frac{\bar{\theta}}{2 \pi}\Delta \phi.\label{precision_initial}
	\end{align}
	This shows the ultimate limit of angular precision of current single-baseline method in principle, and indeed it is independent of baseline length $X$. In order to achieve the exponential angular precision in uncertainty $\Delta\theta$, the current single-baseline method requests the phase uncertainty $\Delta \phi$ exponentially small. For any baseline length $X$, any phase $\phi$, any detection strategy (e.g., observing the values of $\cos\phi$ and $\sin\phi$) and even though experimental set-up itself is perfect, exponentially small  value of error$\Delta \phi$ in single-baseline method requests as least exponentially large number $N$ of detected photons, because of $\Delta \phi\gtrsim\frac{1}{\sqrt{N}}$ given by Cramer-Rao lower bound \cite{huang2022imaging} (detailed calculation is shown in Fig. \ref{epsilon_precision}). We see the current single-baseline method is very difficult to reach a high angular precision in practice since photons are the most precious source in astronomical detection.
	
	To solve this bottleneck problem, here we propose a robust method named in piecemeal array using experimental data from $K$ baselines with length $L_k=2^{k-1}L_1$ ($\{k=1,2,...,K\}$). According to Eq. (\ref{precision}), our piecemeal array improves precision exponentially through
	\begin{align*}
		\Delta\theta=\frac{\lambda}{4L_K}=\frac{\bar{\theta}}{2^{K+1}}.
	\end{align*}
	Different from Eq. (\ref{precision_initial}) of single-baseline method, our piecemeal array benefits from large baseline length $L_K$ and so the uncertainty is exponentially reduced from $\bar\theta$ to $\bar{\theta}/2^{K+1}$. More clearly, we introduce a compression factor for the uncertainty of the phase angle $\theta$ in each method: namely, the ratio of the angle uncertainty after observation to that before observation, expressed as $\Delta\theta/\bar\theta$. In our method, $\Delta\theta/\bar\theta=2^{-(K+1)}$, whereas in the current single-baseline method, $\Delta\theta/\bar\theta=\Delta\phi$. This clearly highlights the precision advantage of our method. Our method only needs the same technologies of the single-baseline experiment, because our piecemeal array only needs to observe the phase difference value $\phi_k$ on each individual baseline. Importantly, to obtain the exponential precision above in our method, we only need crude information of phase difference $\phi_k$ respecting for each individual baseline $L_k$ rather than highly precise information, e.g., an error less than $\pi/3$ of each baseline is enough as shown in Eq. (\ref{e_condition}). Below we present our piecemeal array (Sec. III) and its advantages of exponential precision (Conclusion 1), strong robustness (Sec. IV, Conclusion 2) and high efficiency (Sec. V, Conclusion 2).

	\section*{III. Piecemeal Array}
	
\begin{figure}[htbp]
		\begin{minipage}{0.8\linewidth}
			\includegraphics[width=1\linewidth]{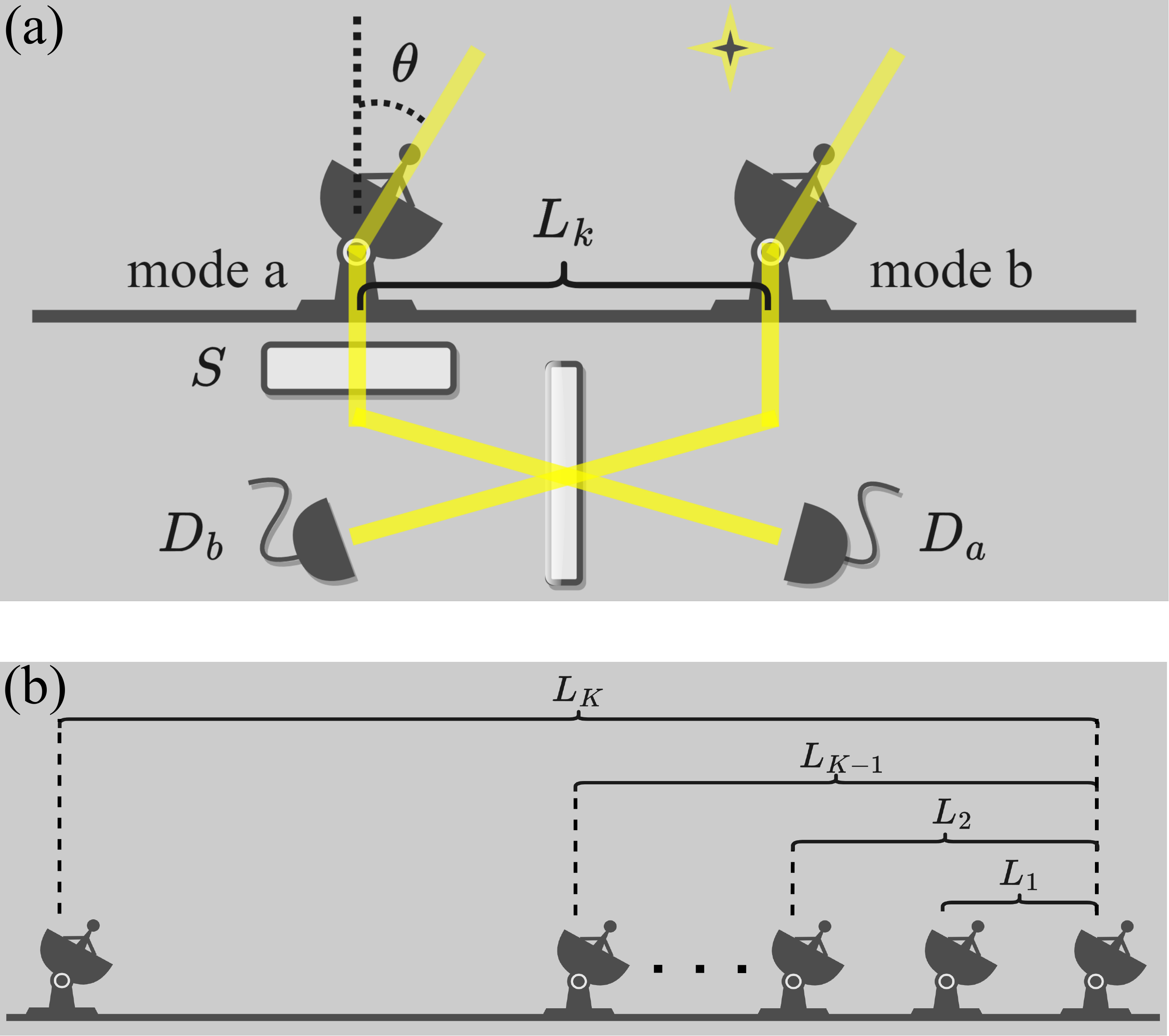}
		\end{minipage}
		\caption{\label{protocol_1}Schematic setup of our method: it contains different baselines $\{L_k\}$. (a) Collecting data from a pair of receivers (telescopes) with phase shifter $S$, 50:50 beam splitter, threshold detectors $D_a$ and $D_b$. We collect the first (second) set of data through setting zero ($\frac{\pi}{2}$) phase shift on $S$. (b) The single-photon interference experiment on each individual baselines with a pair of telescopes are performed separately. One collects data from each individual baseline $L_k$, and then processes the data jointly by Eq. (\ref{piecemeal_final}). Our method requests different baseline lengths only. A realization is not limited to schematic subfigure (b) above. For example, each baselines do not have to be on the same line, they can be on different parallel lines.}
	\end{figure}

	The set-up of our piecemeal array is schematically shown in Fig. \ref{protocol_1}. There are $K$ baselines and we collect data from each of them. The length of baseline $k$ is $L_k$, where $k=\{1,...,K\}$. For ease of presentation, here we take simple relation of
	$L_k = 2^{k-1} L_1$. Suppose that the angle of remote object is $\theta$.
	Suppose the light arriving at two receivers of pair $k$ attenuates to single-photon whose state is
	\begin{align}
		|\mathcal{S}_k\rangle=\frac{1}{\sqrt{2}}(|01\rangle+e^{i\phi_k}|10\rangle),~~\phi_k\approx2\pi\frac{L_k}{\lambda}\theta.\label{single_photon}
	\end{align}
	Here $\lambda$ is the wavelength, $|c_ac_b\rangle$ represents photon numbers of mode $a$ and $b$ and $\phi_1\in[0,~2\pi)$. We start with some mathematical notations.\\
	1) Taking $L_k=2^{k-1}L_1$ and hence $\phi_{k+1} = 2\phi_k$.\\
	2) $\mathring{x} = x$ mod $2\pi$. When using $\alpha \cdot \mathring{x}$, we mean $\alpha \cdot (x\bmod 2\pi)$.\\
	3) $\hat{x}_k$: The observed value of $\mathring{x}_k$. 
	
	In the observation of each baseline, say baseline $k$, we shall have an observed value $\mathring{\phi}_k$ (The detailed calculation formula for $\mathring{\phi}_k$ based on experimental data is shown in Sec. V). Most importantly, we shall treat the observation error $e_k$ in $\mathring{\phi}_k$. A very tricky point here is that we cannot use the simple-minded formula of $\mathring{\phi}_k-\hat{ \phi_k}$ to represent the observation error $e_k$. For example, it is possible that the actual phase $\hat{ \phi_k}$ is small positive value, while our observed value $\mathring{\phi}_k$ is rather large, say a little bit smaller than $2\pi$. In such a case, the observation error is actually very small but the value $\mathring{\phi}_k-\hat{ \phi_k}$ is very large. To treat the observation error reasonably, we introduce the novel modulo calculation below. The actual value $\mathring{\phi}_k$ and its observed value $\hat{ \phi_k}$ can be related by
	\begin{align}
		\hat{\phi}_k = (\mathring{\phi}_k + e_k)\bmod{2\pi}, ~~\text{for any } k,
	\end{align}
	where $e_k \in (-\pi, \pi]$ is the observation error. We introduce a new variable
	\begin{align}
		\psi_k = \phi_k-\hat{\phi}_k+\pi,\nonumber
	\end{align}
	and its modulo
	\begin{align}
		\mathring{\psi}_k=(\phi_k-\hat{\phi}_k+\pi)\bmod 2\pi.\nonumber
	\end{align}
	The definition of $\psi_k$ means $\psi_{k+1} =\phi_{k+1}-\hat{\phi}_{k+1}+\pi$.
	
	\textbf{Theorem 1:} Under the condition
	\begin{align}
	\left|e_{k+1}-2 e_{k}\right|<\pi,\label{e_condition}
	\end{align}
	we have
	\begin{align}
		2 \cdot\mathring{\psi}_k = \mathring{\psi}_{k+1}+ \mathring{r}_{k+1},\label{psi_r_Z}
	\end{align}
	where $r_{k+1}=\hat{\phi}_{k+1}-2\hat{\phi}_k+\pi$.
	
	\rightline{$\square$}

	The proof is shown in Sec. IV. Very importantly, Eq. (\ref{psi_r_Z}) can hold with a small failure probability, because the failure case for Eq. (\ref{e_condition}) requests very large observation error $\{e_k,e_{k+1}\}$ in observed data. Details of fault-tolerance property will be studied in next sections. Using Eq. (\ref{psi_r_Z}) iteratively we have 
	\begin{align*}
		\mathring{\psi}_1 = \sum_{k=2}^K (\frac{1}{2})^{k-1}\mathring{r}_k + (\frac{1}{2})^{K-1} \mathring{\psi}_K.
	\end{align*}
	Recalling the definition $\mathring{\psi}_1=(\phi_1-\hat{\phi}_1+\pi)\bmod 2\pi$ and taking the approximation $\mathring{\psi}_K=\pi$, we obtain 
	\begin{align}
		\phi_1=\mathring{\phi}_1=\left(\hat{\phi}_1-\pi+\sum_{k=2}^K (\frac{1}{2})^{k-1}\mathring{r}_k + (\frac{1}{2})^{K-1}\pi\right)\bmod 2\pi,\label{piecemeal_final}
	\end{align}
	where $\mathring{\phi}_1=\phi_1$ has been used since $\phi_1\in[0,~2\pi)$. This is the major equation in this work. And this gives the result of angular value
	\begin{align}
		\theta_1=\frac{\lambda}{2\pi L_1}\phi_1.\label{piecemeal_final_2}
	\end{align}
	Suppose the initial uncertainty of angle is $0<\theta_1<\bar{\theta}$, we obtain $L_1=\lambda/\bar\theta$ by Eq. (\ref{piecemeal_final_2}). In Eq. (\ref{piecemeal_final}), the uncertainty comes from replacing the term $(\frac{1}{2})^{K}\mathring{\psi}_K$ by $(\frac{1}{2})^{K}\pi$. This gives the precision of our piecemeal array by
	\begin{align}
		\Delta\phi=\frac{\pi}{2^{K}},~~\Delta\theta=\frac{\lambda}{2\pi L_1}\Delta\phi=\left\{
		\begin{aligned}
			&\frac{\lambda}{4L_K},\text{ or equivalently,}\\
			&\frac{\bar\theta}{2\pi}\Delta\phi=\frac{\bar\theta}{2^{K+1}}.
		\end{aligned}
		\right.\label{precision}
	\end{align}
	where $L_1=\lambda/\bar\theta$ has been used. The precision in uncertainty is proportional to $(\frac{1}{2})^{K}$, i.e., $K$ bits of value has been determined, which shows a bit-by-bit behavior.
	
	\textbf{Conclusion 1:} Linear scale of resources for exponential precision. Our method assumes $K$ pairs telescopes, while the precision in uncertainty is proportional to $(\frac{1}{2})^K$. This means the precision improves much faster than the number of telescopes and the cost of space resource. In the visible light band ($\lambda\approx10^{-7}$m), the length of current optical interference can reach about $10^3$m \cite{array}. Applying Eq. (\ref{precision}) we obtain the angular precision about $10^{-2}$mas of our piecemeal array. The typical resolution limit of current large telescope is about $10^2$mas \cite{telescopes}. So our method improves the angular precision by 4 magnitude orders comparing with current large telescopes. This improvement indeed comes from the large equivalent diameter of optical interference and the tricky design of piecemeal array, as we already expected in Sec. II.
	
	Remark: In the first option of our piecemeal method, we set all baselines in one straight line and we only need $K+1$ telescopes. In the second option of our piecemeal method we need $2K$ telescopes which can run simultaneously for all telescopes by adding beam splitters after each pair of telescopes. 
	
	\section*{IV. Fault Tolerance Property}
	
	\textbf{Proof of Theorem 1:} 
	Observe the identity in modulo arithmetic:
	\begin{align*}
		(x\bmod{2\pi}  + y)\bmod{2\pi} = (x+y)\bmod{2\pi}.
	\end{align*}
	Applying this identity, we derive the following:
	\begin{align*}
		2\cdot\mathring{\psi}_k = &2[(\mathring{\phi}_k-(\mathring{\phi}_k + e_k)\bmod{2\pi} + \pi)\bmod{2\pi}]\\
		=&2 \pi - 2 e_k,  \\
		\mathring{\psi}_{k+1} = &(\mathring{\phi}_{k+1} - (\mathring{\phi}_{k+1} + e_{k+1})\bmod{2\pi}+\pi)\bmod{2\pi}\\
		=&\pi - e_{k+1},\\
		\mathring{r}_{k+1} = &((\mathring{\phi}_{k+1} + e_{k+1})\bmod{2\pi} - 2 (\mathring{\phi}_k + e_k)\bmod{2\pi} + \pi)\\
		&\bmod{2\pi} \\
		= &((\mathring{\phi}_{k+1} - 2 (\mathring{\phi}_k))\bmod{2\pi} + e_{k+1} - 2 e_k + \pi)\bmod{2\pi},
	\end{align*}
	where Eq. (\ref{e_condition}) has been used. Given $\mathring{\phi}_{k+1} = (2\cdot\mathring{\phi}_k)\bmod{2\pi}$, it follows that: 
	\begin{align*}
		(\mathring{\phi}_{k+1} - 2\cdot\mathring{\phi}_k) \bmod{2\pi} = 0.
	\end{align*}
	Consequently,
	\begin{align*}
		\mathring{r}_{k+1} =  (e_{k+1} - 2 e_k + \pi)\bmod{2\pi} = e_{k+1} - 2 e_k + \pi.
	\end{align*}
	Hence,
	\begin{align*}
		&2\cdot\mathring{\psi}_k - \mathring{\psi}_{k+1} - \mathring{r}_{k+1}\\
		=&2 \pi - 2 e_k - (\pi - e_{k+1}) - (e_{k+1} - 2 e_k + \pi)\\
		=&0.
	\end{align*}
	\rightline{$\square$}
	
	The key point of proof is the condition of Eq. (\ref{e_condition}). We see that the event with $\left|e_{k+1}-2 e_{k}\right|\ge\pi$ is very unlikely because it can only happens with large observation errors $\{e_k,e_{k+1}\}$. The observation error has already included all the noise effects. Therefore Eq. (\ref{e_condition}) means that our method with the calculation formula Eq. (\ref{psi_r_Z}) has a strong fault tolerance property to any noise effect, surely including the large channel noise (e.g., atmospheric disturbance) and the statistical fluctuation caused by small number of incident single-photons.
	
	As a simple example, a sufficient condition for Eq. (\ref{e_condition}) is that the observed phase value resides in the same quadrant with the exact phase value for $k$ and $k+1$. (There are four quadrants in $[0,~2\pi)$.)
	
	\textbf{Fact 1:} Our method works exactly provided that the observed value $\hat{\phi}_k$ is in the same quadrant with the exact value $\mathring{\phi}_k$ for all $k$.
	
	To shown Fact 1, we use the following conditions jointly: 1) $\mathring{\phi}_k$ and $\hat{\phi}_k$ are in the same quadrant; 2) $\mathring{\phi}_{k+1}$ and $\hat{\phi}_{k+1}$ are in the same quadrant; 3) $\mathring{\phi}_{k+1}=(2\phi_k)\bmod2\pi$. We simply consider eight cases below one by one: \{$\hat{\phi}_k$, $\mathring{\phi}_k$\} are in quadrant $\alpha$ while \{$\hat{\phi}_{k+1}$, $\mathring{\phi}_{k+1}=(2\mathring{\phi}_k)\bmod2\pi$\} are in quadrant $[2(\alpha -1) +1]\bmod 4$ or quadrant $[2(\alpha -1) +2]\bmod 4$, where $\alpha=\{1,~2,~3,~4\}$. Fact 1 is correct in all these eight cases.
	
	Consider the data collection with $L_k$ in Fig. \ref{protocol_1}(a) in the asymptotic case. Suppose in the noiseless case, the first data set is $(n_a,~n_b)$ and the second data set is $(n_a^{\prime},~ n_b^{\prime})$, where $n_a$ ($n_b$) and $n_a^{\prime}$ ($n_b^{\prime}$) are the number of counts at detector $D_a$ ($D_b$) in each data set. Now consider the noise case with flipping error model. The observed values of counts at each detectors of the first data set become to:
	\begin{align}
		\begin{cases}
			\hat{n}_a = n_a (1 - p_a) + n_b p_b\\
			\hat{n}_b = n_b (1 - p_b) + n_a p_a.\label{equ_4_15}
		\end{cases}
	\end{align}
	where $p_a$, $p_b$ are flipping rates, and $(\hat{n}_a,~\hat{n}_b)$ are the observed values of counts at each detectors of the first data set. Similar equation also holds for the second data set:
	\begin{align}
		\begin{cases}
			\hat{n}^{\prime}_a = n^{\prime}_a (1 - p^{\prime}_a) + n^{\prime}_b p^{\prime}_b\\
			\hat{n}^{\prime}_b = n^{\prime}_b (1 - p^{\prime}_b) + n^{\prime}_a p^{\prime}_a.\label{equ_4_16}
		\end{cases}
	\end{align}
	where $p^{\prime}_a$, $p^{\prime}_b$ are flipping rates. With $n_a,n_b,n_a^{\prime},n_b^{\prime}$ ($\hat{n}_a,\hat{n}_b,\hat{n}^{\prime}_a,\hat{n}^{\prime}_b$), we can determine the value of $\{\mathring{\phi}_k\}$ ($\{\hat{\phi}_k\}$) by Eq. (\ref{data_phi}). Here we only need to know the specific quadrant of $\{\hat{\phi}_k\}$ and $\{\mathring{\phi}_k\}$, which are determined by the following fact:
	
	\textbf{Fact 2:} For any $k$, the quadrant of $\mathring{\phi}_k$ is determined by signs of $n_b - n_a$ and $n_b^{\prime}-n_a^{\prime}$, and the quadrant of $\hat{\phi}_k$ is determined by $\hat{n}_b - \hat{n}_a$ and $\hat{n}_b^{\prime}-\hat{n}_a^{\prime}$.
	
	It is to say that, the flipping error model given by Eq. (\ref{equ_4_15}) and Eq. (\ref{equ_4_16}) does not change the sign of any difference term $x-y$ as appeared in Fact 2 above if all flipping rates are less than $50\%$.
	Therefore the observed value $\hat{\phi}_k$ ($\hat{\phi}_{k+1}$) and the exact value $\mathring{\phi}_k$ ($\mathring{\phi}_{k+1}$) are in the same quadrant provided that all flipping rates are less than $50\%$. We conclude:
	
	\textbf{Theorem 2:} Asymptotically, our piecemeal array with Eq. (\ref{piecemeal_final}) work exactly with flipping errors if the flipping rates are less than $50\%$. 
	
	As a direct application we have the following theorem:
	
	\textbf{Theorem 3:} Asymptotically, our piecemeal array with Eq. (\ref{piecemeal_final}) work exactly under \textbf{any noisy channel} of random phase drift $\delta\phi_k$ in the range of $(-\pi,~\pi)$ for setting zero phase shift of $S$ and $(-\pi,~\pi)$ for setting $\frac{\pi}{2}$ phase shift of $S$.
	
	Theorem 2 and 3 actually show that our method has strong robustness to large channel noise. In the next section, we consider more general case of large channel noise and small number of single-photons.

	\section*{V. Performance under channel noise and Small number of single-photons}
	
	In the proposed set-up Fig. \ref{protocol_1}, we collect two sets of data. The first (second) set is obtained through setting zero ($\frac{\pi}{2}$) phase shift on $S$. In the first (second) set, we observe the event of detector $D_a$ silent and $D_b$ clicking for $m_k$ ($\overline{m}_k$) times in total detected events of one-detector-clicking $M_k$ ($\overline{M}_k$). We have
	\begin{align}
		\hat{\phi}_k=\frac{1}{2}\left(\cos^{-1}q_k+\sin^{-1}\overline{q}_k\right)\bmod 2\pi,\label{data_phi}
	\end{align}
	where $q_k=1-\frac{2m_k}{M_k}$ and $\overline{q}_k=\frac{2\overline{m}_k}{\overline{M}_k}-1$. Since $M_k$ and $\overline{M}_k$ are finite number in practical detection, the $\phi_1$ value calculated by Eq. (\ref{piecemeal_final}) does not exactly equal to the actual value of $\phi_1$. Denoting $\tilde{\phi}_1$ ($\tilde{\theta}$) as the $\phi_1$ ($\theta$) value calculated by Eq. (\ref{piecemeal_final}) (Eq. (\ref{piecemeal_final_2})), we define the failure probability $\epsilon$ as
	\begin{align}
		\epsilon&=P(|\tilde{\phi}_1-\phi_1|>\Delta\phi=\frac{\pi}{2^{K}})\nonumber\\
		&=P(|\tilde{\theta}-\theta|>\Delta\theta=\frac{\lambda}{4L_K}).\label{epsilon}
	\end{align}
	Taking average over $\phi_1\in[0,~2\pi)$ we obtain the average failure probability $\langle\epsilon\rangle$. Below we calculate the relationship of average failure probability $\langle\epsilon\rangle$ and total single-photon number $N$ by two steps. To quantify the performance, we set $\lambda=380$nm, initial uncertainty of angle $0\le\theta<1.2\text{as}$, largest baseline length $L_K=1.07$km and $M_k=\overline{M}_k=M$ for all $k$. We obtain $L_1=\lambda/1.2\text{as}=0.065$m from Eq. (\ref{piecemeal_final_2}), $K=\log_2(L_K/L_1)+1=15$, phase precision in uncertainty $\Delta\phi =9.59\times10^{-5} $ and angular precision in uncertainty $\Delta\theta=0.0183$mas in Eq. (\ref{epsilon}).
	
	Step 1, calculate the relationship of $\langle\epsilon\rangle$ and $M$. Given single-photon state Eq. (\ref{single_photon}), the probabilities of detector $D_b$ or $D_a$ clicking is $p_k=\frac{1}{2}(1-\cos(\phi_k+\delta\phi_k))$ and $\overline{p}_k=\frac{1}{2}(1+\sin2^k(\phi_k+\delta\phi_k))$, respectively. Here the channel noise $\delta\phi_k$ obeys the Gaussion distribution $\mathcal{N}(0, \sigma^2)$. The atmospheric effect is already included in channel noise. With $M$ and $p_k$ ($\overline{M}_k$ and $\overline{p}_k$), we obtain the value of $m_k$ ($\overline{m}_k$) by randomizer. Using $\{m_k,\overline{m}_k\}$, we calculate the estimated value $\tilde{\phi}_1$ by Eq. (\ref{piecemeal_final}). If $|\tilde{\phi}_1-\phi_1|<\Delta \phi$, we regard it as a successful event otherwise a failure event. We see that the average failure probability $\langle\epsilon\rangle$ only depends on $M$ and noise parameter $\sigma$. 
	
	Step 2, calculate the relationship of $N$ and $M$. The channel transmittance is $\eta(l)$ of optical fiber of length $l$. Suppose that beam splitter locates at the middle of fiber. So the incident single-photons at receives have a probability $\eta(\frac{L_k}{2})$ to arrive at detectors. We obtain the total number of collected single-photons at receivers $N=\sum_{k=1}^{15}\frac{2M}{\eta(L_k/2)}$. We set $\eta(l)=10^{(-\alpha l / l_0)}$ where $\alpha=0.2$ and $l_0=10$km, so the total single-photon number $N$ only depends on $M$.
	
\begin{figure}[htbp]
		\begin{minipage}{1\linewidth}
			\includegraphics[width=0.9\linewidth]{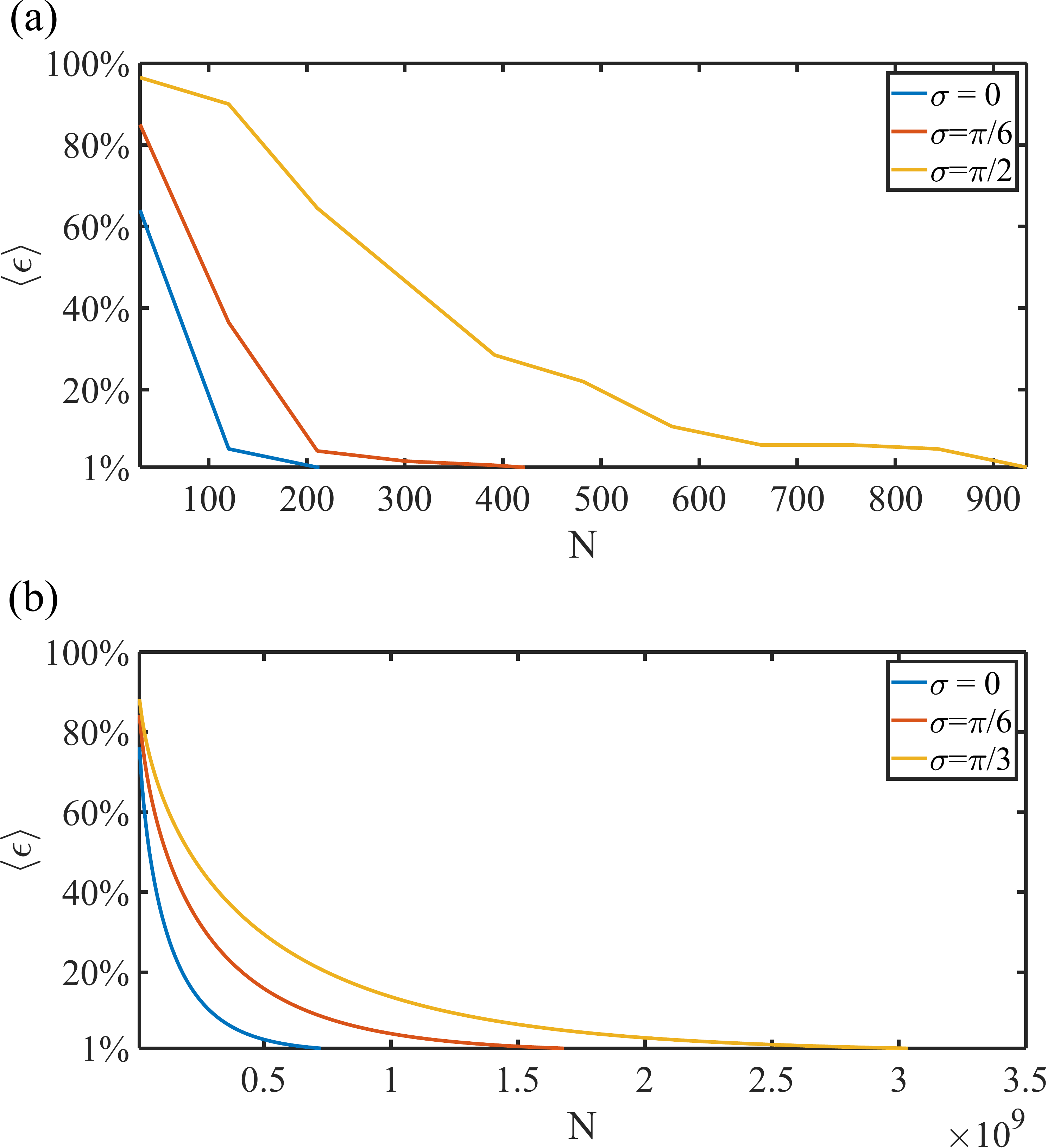}
		\end{minipage}
		\caption{\label{epsilon_precision}The average failure probability $\langle\epsilon\rangle$ in Eq. (\ref{epsilon}) changes with the total single-photon number $N$, under the Gaussian distribution noise $\mathcal{N}(0, \sigma^2)$. We set largest baseline length $L_K=1.07$km, $\lambda=380$nm, initial uncertainty of angle $\theta\in[0,~1.2\text{as})$ and the angular precision in uncertainty $\Delta\theta=0.0183$mas. (a) Our piecemeal array. (b) Current single-baseline method \cite{gottesman2012longer,khabiboulline2019optical,huang2022imaging,marchese2023large}. To reach the precision in uncertainty of $0.0183$mas with failure probability lower than $1\%$, our method requires $10^2\sim10^3$ single-photons, while single-baseline method requires $10^{9}$ single-photons.}
	\end{figure}

	After these two steps, we obtain the relationship of average failure probability $\langle\epsilon\rangle$ and total single-photon number $N$ of our piecemeal array, as shown in Fig. \ref{epsilon_precision}(a).
	Applying 3-$\sigma$ principle, we see that our piecemeal array works well in the large noise range with $[-3\sigma,3\sigma]=[\pi,\pi]$ (equivalent to $[-\lambda/2,\lambda/2]$ in the language of wavelength). As comparison, we calculate the relationship of $\langle\epsilon\rangle$ and $N$ for current single-baseline method \cite{gottesman2012longer,khabiboulline2019optical,huang2022imaging,marchese2023large}, using the same initial uncertainty $0\le\theta<\text{1.2as}$. Applying the Central Limit Theorem, the distribution of estimated angular value by single-baseline method obeys Gaussian distribution with the standard deviation $\sigma_\theta\ge\frac{\lambda(1+\sigma)}{2\pi L_1 \sqrt{N}}=\frac{(1+\sigma)\text{1.2as}}{2\pi \sqrt{N}}$ \cite{huang2022imaging}, where $\sigma$ is the standard deviation of channel noise $\mathcal{N}(0, \sigma^2)$ and $L_1=\lambda/1.2\text{as}$ has been used. We take the lower bound of $\sigma_\theta$ can calculate the average failure probability by Gaussian Cumulative Distribution Function $\langle\epsilon\rangle=\langle 2F(-\Delta\theta,\sigma_\theta )\rangle=2F(-\Delta\theta,\sigma_\theta)$. The relationship of $\langle\epsilon\rangle$ and $N$ for single-baseline method is shown in Fig. \ref{epsilon_precision}(b).
	
	\textbf{Conclusion 2:} Strong robustness and high efficiency. With large noise range of $[-\pi,\pi]$ and requiring only $10^2\sim10^3$ single-photons, our piecemeal array can reach the precision in uncertainty of 0.0183mas with negligible failure probability. As a comparison, with same noise channel, same precision and same failure probability, the current single-baseline methods require $10^{9}$ single-photons. So our method reduces the requirement of single-photon number by $6\sim7$ magnitude orders comparing with single-baseline methods. This resource reduction indeed comes from the strong fault tolerance our method which is robust to any noise effect including the large channel noise and the statistical fluctuation caused by small number of single-photons, as we already proved in Sec. IV.
	
	\section*{VI. Piecemeal Array with Reference object for astronomical detection}
	
	Our method can be applied to detect the angle between target object and reference object for astronomy. Since these two objects are close, either the complicated atmospheric disturbance or the fluctuation of long baseline causes almost the same effect on the phase values of target and reference objects. Our method can work robustly when applied to detect the angle between the target object and the reference object.
	
	Consider two remote objects: the reference R and the target T. Our goal is to detect the angle of T relative to R, i.e., angle $\Gamma$ in Fig. \ref{fig_7_1}. The value of $\Gamma$ can be around the magnitude order of 1 arcsecond. Suppose we have some crude information about angle $\Gamma$ and we write it in the following form:
	\begin{figure}[htbp]
		\includegraphics[width=0.4\linewidth]{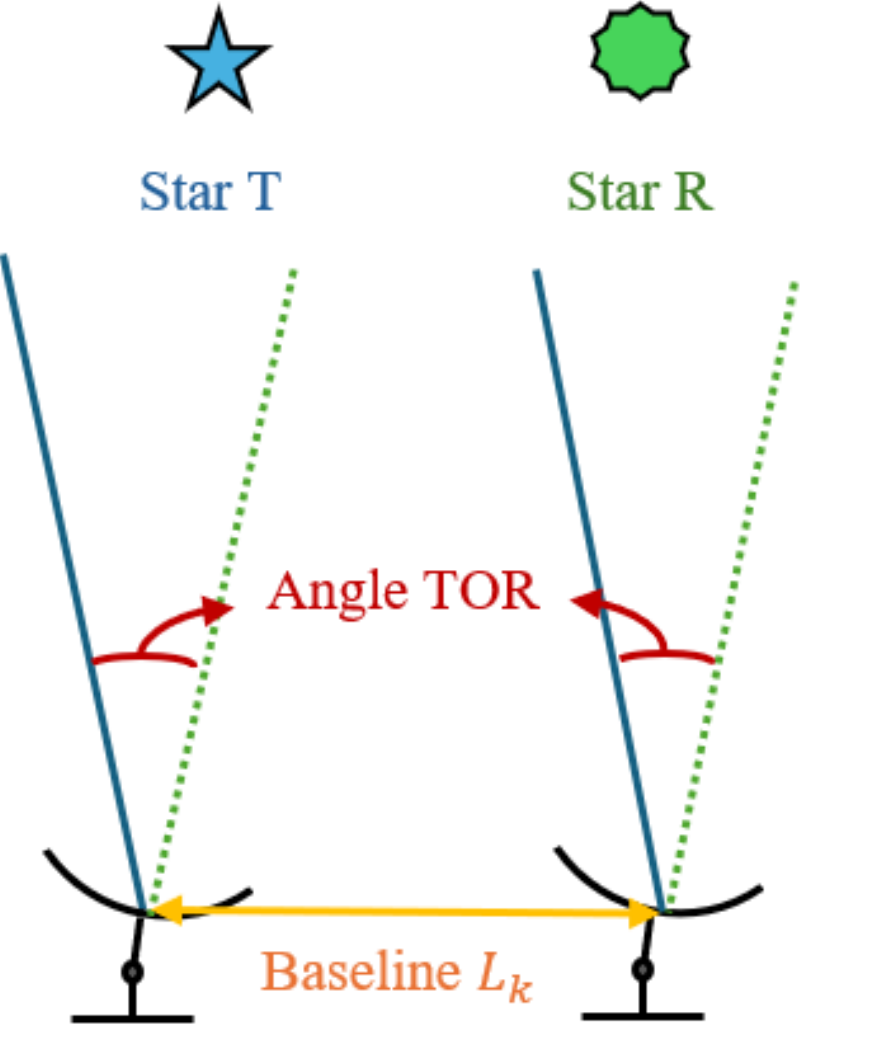}
		\caption{\label{fig_7_1}Using a reference star with a known position can help reduce the failure probability inherent to the experimental devices. For each baseline length $L_k$, photons from both the target T and the reference R are received and the experimental data are processed jointly, as shown in Eq. (\ref{equ_7_2}).}
	\end{figure}
	\begin{align*}
		\Gamma = \Gamma_0 + \theta_T,
	\end{align*}
	where $\Gamma_0$ is an exact value known to us while $\theta_T$ is an unknown small value whose range is $\theta_T \in [0,\theta_0]$, value $\theta_0$ is known precisely. The value of $\Gamma_0$ can be around 1 arcsecond or several arcseconds, while the value of $\theta_0$ can be around magnitude of 1 milliarcsecond. Therefore we can distinguish incident photons from different stars and hence can perform the interference experiment for photons from different stars separately.
	We denote $\Phi_{T,k}$ and $\Phi_{R,k}$ for phase difference of incident light to telescopes at baseline $k$ from star T and R, respectively. In particular, 
	\begin{align*}
		\Phi_{R,k} =& \frac{2\pi L_k}{\lambda} \sin \theta_R\nonumber\\
		\Phi_{T,k}=& \frac{2\pi L_k}{\lambda} \sin \left(\Gamma + \theta_R\right)\nonumber\\
		=&\frac{2\pi L_k}{\lambda} \sin \left(\Gamma_0+\theta_T + \theta_R\right),
	\end{align*}
	where $\theta_R$ (or $\Gamma + \theta_R$) is the angle between baseline $k$ and wave plane of light from star R (or wave plane of light from star T).
	
	In our setup, the baselines are supposed to be (almost) parallel to each other and parallel to plane TOR. Here O is any point around the telescopes. (If the sources are remote stars T and R, we can simply regard the earth as point O.) The baselines are also supposed to be (almost) orthogonal to wave ray of light from star R. This means $\theta_R$ $\sim$ 0. The difference of phase difference is
	\begin{align*}
		\Phi_{T,k}-\Phi_{R,k} = \frac{2\pi L_k \sin \Gamma}{\lambda}
	\end{align*}
	Taking the conditions that $\theta_R$ $\sim$ 0, $\Gamma_0$ around arcseconds, we have
	\begin{align*}
		\Phi_{T,k}-\Phi_{R,k} \approx \frac{2\pi L_k}{\lambda} (\sin \Gamma_0 + \cos \Gamma_0 \sin \theta_T).
	\end{align*}
	Denoting 
	\begin{align}\label{equ_7_1}
		\frac{2\pi L_k}{\lambda} \cos \Gamma_0 \sin \theta_T = \Delta_k,
	\end{align}
	we have
	\begin{align*}
		\Delta_k = \Phi_{T,k}-\Phi_{R,k} - \frac{2\pi L_k}{\lambda} \sin \Gamma_0.
	\end{align*}
	Denoting $\Psi_k =\Delta_k - \hat{\Delta}_k + \pi$, $\mathring{\gamma}_{k+1} = (\hat{\Delta}_{k+1} - 2\hat{\Delta}_{k} + \pi)\mod 2\pi$ and taking the iteration and derivation similar to those used for Eq. (\ref{psi_r_Z}), we obtain
	\begin{align}
		\mathring{\Delta}_1=(\hat{\Delta}_1-\pi + \sum_{k=2}^{K}(\frac{1}{2})^{k-1}\mathring{\gamma}_k+(\frac{1}{2})^{K-1}\pi)\mod 2\pi,\label{equ_7_2}
	\end{align}
	which can be observed experimentally because values of terms at the right side of the Eq. (\ref{equ_7_2}) are either experimentally observable or exactly known.	It is just $\Delta_1$ if we set $L_1 \sin \theta_1 < 2\pi$. By Eq. (\ref{equ_7_2}), we can calculate $\theta_T$ according to Eq. (\ref{equ_7_1}) with $k=1$.
	
	\section*{VII. Discussion}
	
	We propose a new detection method through bit-by-bit iteration, as an enhanced technique for optical telescopes. Our method requests only small number of single-photons, holds strong fault tolerance to any noise and improves the precision by magnitude orders comparing with current optical telescopes. Given these advantages, our method promises a progress in remote sensing and astrometry. In remote sensing, our method can be applied to detect the optical information of very dark object in the night. In such a case, regular optical detection methods are always invalid. In astrometry, our method can not only improve the angular precision of stars detected by present large optical telescopes, but also detect the very dark stars that have not been discovered so far. For example, detecting the orbit of very dark stars around the black hole. This is quite meaningful since the photons emitted by stars are very precious resources in astronomical detection \cite{malbet2021faint,baker2021high,vallenari2018future,rioja2020precise}.
	
	We for the first time propose the theoretical scheme of piecemeal array and also consider the most important practical conditions of small photon number, channel noise (including atmospheric effect) and fiber attenuation. As reported in Ref. \cite{shen2025piecemeal}, our piecemeal method is also strongly robust to the set-up imperfections such as the baseline length error, the baseline orientation error and so on.
	
	\begin{acknowledgments}
		We acknowledge the financial support in part by National Natural Science Foundation of China grant No.11974204 and No.12174215, and Innovation Program for Quantum Science and Technology No. 2021ZD0300705. This study is also supported by the Taishan Scholars Program. We thank Prof. Y Cao of USTC for useful discussions.
	\end{acknowledgments}
	
	\bibliography{refs}
	
\end{document}